\documentclass[aps,showpacs,groupedaddress]{revtex4}
\pdfoutput=1

\usepackage{graphicx}

\newcommand{\micron}{\ensuremath{\mu\mbox{m}}}

\begin{document}

\title{Observation of simultaneous fast and slow light}

\author{Pablo Bianucci}
  \email{bianucci@ualberta.ca}
  \altaffiliation{Now at Department of Physics, University of Alberta,
		Edmonton, AB, T6G 2G7, Canada.}
\author{Chris R. Fietz}
\author{John W. Robertson}
\author{Gennady Shvets}
\author{Chih-Kang Shih}
\affiliation{Department of Physics, The University of Texas at Austin, Austin,
	     TX 78712}

\date{\today}

\begin{abstract}
We present a microresonator-based system capable of simultaneously producing
time-advanced and time-delayed pulses.  The effect is based on the combination
of a sharp spectral feature with two orthogonally-polarized propagating
waveguide modes.  We include an experimental proof-of-concept implementation
using a silica microsphere coupled to a tapered optical fiber and use a
time-domain picture to interpret the observed delays.  We also discuss potential
applications for future all-optical networks.
\end{abstract}

% PACS numbers.

\maketitle

% Introduction

Increasing bandwidth demands are pushing for the development of all-optical
circuitry that will be able to quickly and reliably transmit and process vast
amounts of data.  Precise control of light propagation, such as the ability to
advance, delay, or store a pulse transmitted through a waveguide, is a requisite
ingredient for optical data processing in photonic circuits\cite{ChangPIEEE06}. 
Remarkably, exotic optical phenomena occurring in the presence of strong
spectral dispersion such as slow\cite{HauNAT99} and fast\cite{WangNature00}
light have been found very useful for achieving these goals, generating a surge
of experimental activities aimed at realizing fast or slow light in different
media:  atomic
vapors\cite{HauNAT99,WangNature00,CamachoPRL07}, crystals\cite{BigelowSCI03},
semiconductors\cite{Chang-HasnainJLT06,SarkarOPE06}, and
microresonators\cite{VlasovNAT05,XiaNPH06,TotsukaPRE07,TotsukaPRL07}. These
demonstrations have shown either fast or slow light for a given configuration. 
Here we introduce a microresonator-based system that is capable of
simultaneously producing time-advanced and time-delayed pulses, including an
experimental proof-of-concept implementation.

The ability to simultaneously slow and advance pulses of light brings about a
new perspective on photonics:  one can easily envision applications involving
all-optical processing of data headers and data packets where both fast and slow
light may be desirable.  Time-advanced signals can be used to compensate time
delays inevitable in any complex optical-processing network\cite{SolliPRE02}. 
The appeal of strong spectral dispersion is not limited to the linear properties
of light:  the resulting high optical energy compression may lead to
extraordinary enhancement of nonlinear effects and, one day, to single-photon
applications for quantum computing and communications.  Our experimental
implementation uses a silica microsphere evanescently coupled to a tapered
optical fiber\cite{KnightOPL97,CaiPRL00} as the resonator.  Light polarization
plays an important role in the experiment as the enabling tool for achieving
fast/slow light and also provides an additional degree of freedom available for
data storage and/or processing.

It has been known for a long time that media with sharp spectral features can
modify the group speed of light propagating through them\cite{BoydPIP02},
resulting in subluminal or superluminal propagation of pulses.  More recently it
has also been established that systems where two modes can propagate can also
show anomalous dispersion, even in the absence of absorption or
reflection\cite{SolliPRL03}.  This has been shown in photonic
crystals\cite{SolliPRL03,SolliPRL04} and coupled mode
systems\cite{MelloniPRL07}.  Our system combines both approaches, using a
coherent linear superposition of two propagating modes where only one of them
shows a sharp spectral feature\cite{BianucciOPL07,FietzOPL07}.  This enables us
to produce, out of a single incident pulse, two mutually
orthogonal pulses of comparable intensity:  one time-delayed and another
time-advanced with respect to the incident pulse.

% Theory

\begin{figure}[htbp]
  \centering
  \includegraphics[width=8.2cm,keepaspectratio]{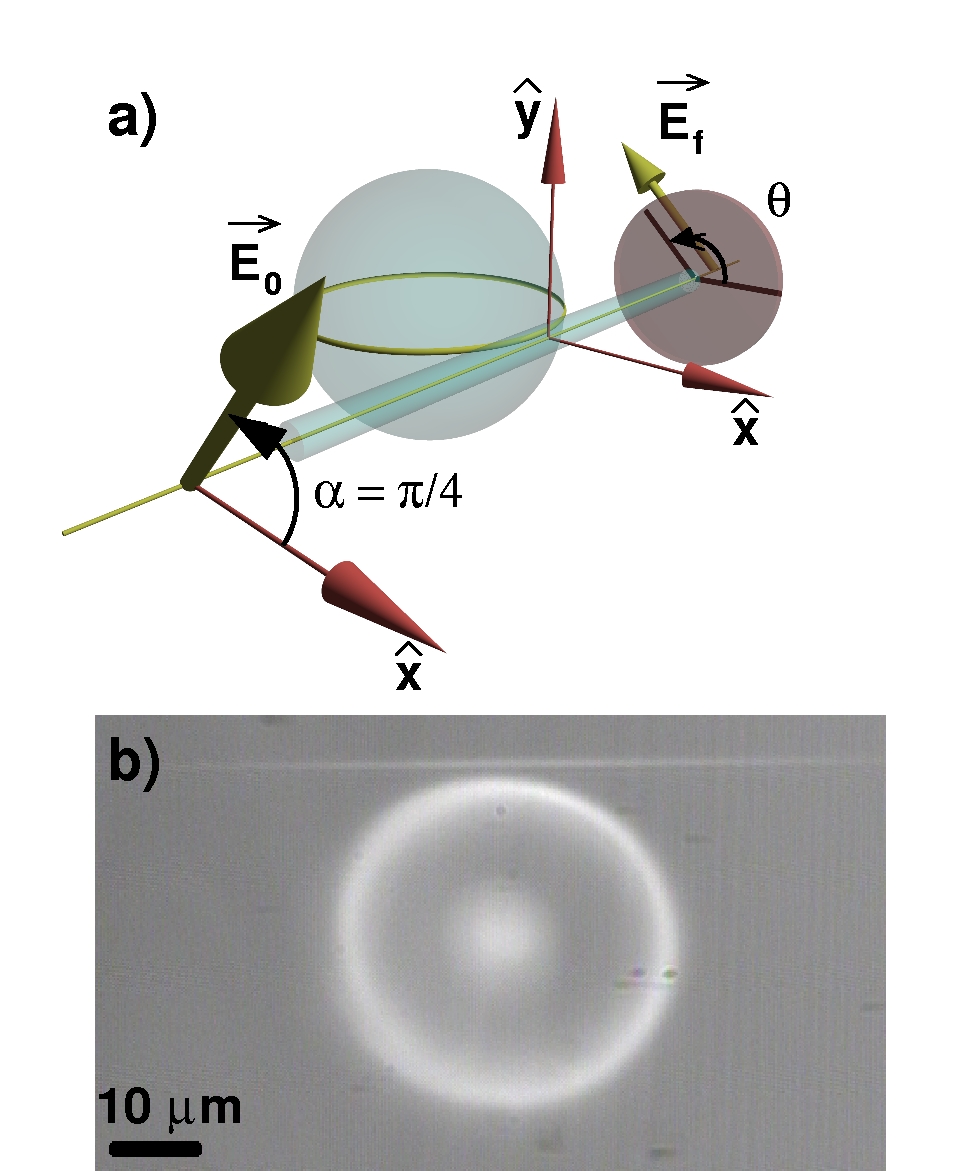}
  \caption{(a) (Color online) The waveguide supports two degenerate modes with
       orthogonal polarizations, coincident with the resonator natural
       polarization axis ($\hat{x}$ and $\hat{y}$).  The incident field
       $\vec{E}_0$ polarization is oriented halfway between the two axis.  A
       polarizer at the output selects either the polarization parallel to the
       incoming one ($\theta=\pi/4$) or the perpendicular one ($\theta=-\pi/4$). 
       (b) Optical microscope picture showing a microsphere close to a tapered
       optical fiber.}
  \label{fig:schem}
\end{figure}

We consider a resonator (microsphere) strongly coupled to a waveguide (optical
fiber) and assume that the incoming light is incident with a polarization at
$45$ degrees from the resonator's natural polarization axis, as indicated in
Fig.~\ref{fig:schem}.  An analyzer selects the polarization component parallel
to the incoming one ($\theta=\pi/4$) or orthogonal to it ($\theta=-\pi/4$). 
Assuming that only one of the natural polarizations of the fiber-microsphere
system exhibits resonant transmission features, the intensity transmitted
through the waveguide can be written for each of the output polarizations
as\cite{BianucciOPL07}
\begin{equation}
  I(\pm\pi/4) = \frac{I_0}{4} \left( 1 + |\tau|^2
                                 \pm 2 |\tau|\cos\phi \right),
\end{equation}
\begin{equation}
  \tau = |\tau|e^{i\phi} = \frac{r-ae^{i\varphi}}{1-rae^{i\varphi}},
\label{eq:tau}
\end{equation}
where $a$ and $\varphi$ are the single-pass resonator attenuation and
phase-shift, respectively, and $r^2 = 1 - t^2$ with $t$ representing the
coupling coefficient between the waveguide and the resonator.

\begin{figure}[htbp]
  \centering
  \includegraphics[width=8.2cm,keepaspectratio]{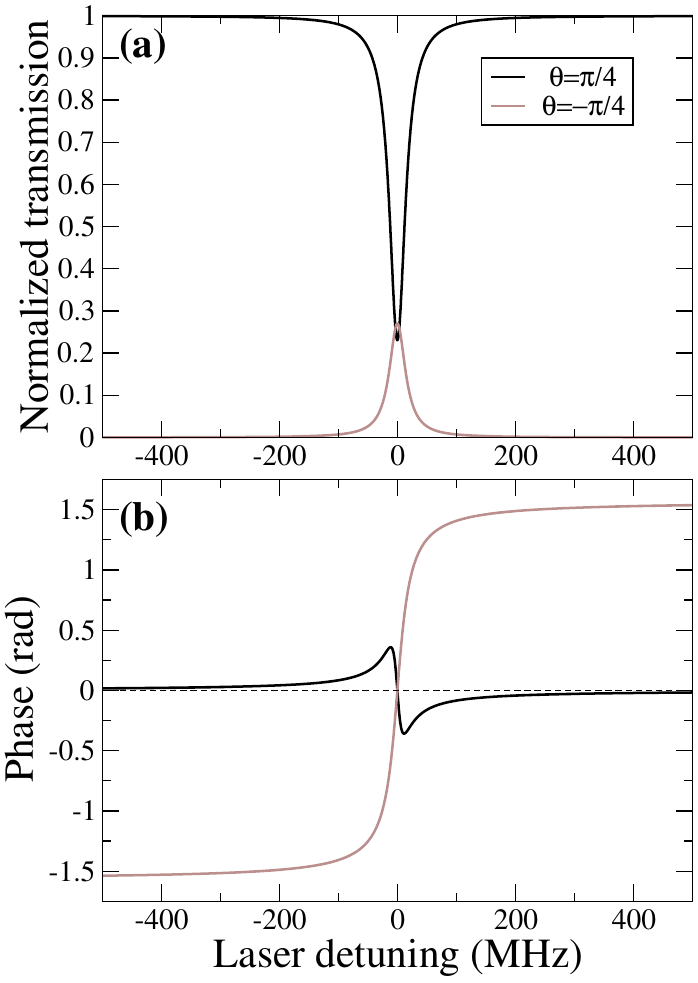}
  \caption{(a) Transmission spectrum for both detected polarizations in a
       slightly overcoupled regime ($a=0.999963, r=0.999960$).  At resonance, a
       fraction of the radiation has its polarization rotated by 90 degrees. 
       (b) Phase spectrum for both polarizations.  The parallel
       polarization shows anomalous dispersion near the resonance, while the
       perpendicular one shows normal but steep dispersion.}
  \label{fig:theo_pol}
\end{figure}

Calculated transmission and phase spectra for each one of the output
polarizations are shown in Fig.~\ref{fig:theo_pol} for the case of a slightly
overcoupled resonator.  The parallel polarization shows clear absorption and
anomalous dispersion at resonance, while the perpendicular one shows gain and
normal, but steep, dispersion.  Further analysis using vectorial Kramers-Kronig
relations\cite{SolliPRL03} suffices to realize that the light coming out with
the analyzer set at $\theta=\pi/4$ will show a negative group delay, while there
will be a positive group delay for the orthogonal polarization.  Using a
polarization beamsplitter instead of an analyzer the incoming pulse can be split
into two, with one of the child pulses advanced in time and the other delayed.

\begin{figure}[htbp]
  \centering
  \includegraphics[width=8.2cm,keepaspectratio]{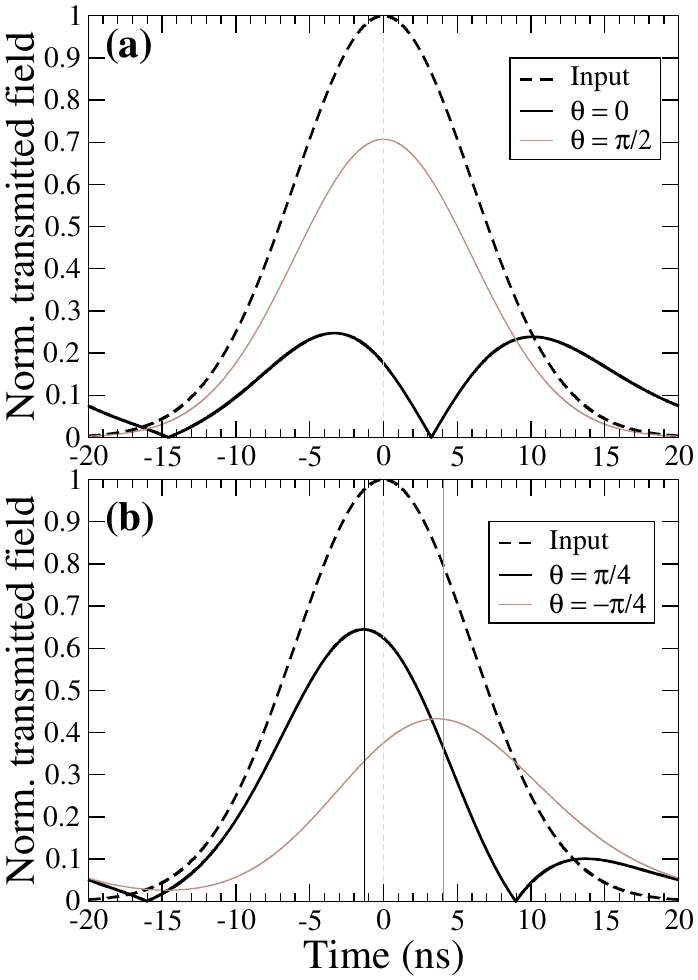}
  \caption{(a) Calculated transmitted pulses in the resonator natural
       polarization axes, for the same conditions as in Fig.~\ref{fig:theo_pol}. 
       The $\theta=\pi/2$ component is affected by the resonator, while the
       $\theta=0$ one is not.  The input pulse is polarized along
       $\theta=\pi/4$.  (b) Calculated transmitted pulses for the parallel and
       orthogonal polarizations.  The peak of the parallel pulse is advanced in
       time, while that of the perpendicular pulse is delayed.}
  \label{fig:theo_time}
\end{figure}

Previous theoretical work\cite{FietzOPL07} used the frequency-domain approach
outlined above to indicate positive and negative group delays are possible for
transmission through a waveguide coupled to a resonator, but this is not the
only possible approach.  The modification of the pulse arrival time can also be
explained from a time-domain perspective that matches quite well the time-domain
nature of the experimental implementation.  In this view, one of the pulse's
polarization components is coupled to the resonator (and is thus subjected to
dispersion coming from the frequency-dependent transmission coefficient in
Eq.~\ref{eq:tau}) while the other travels straight through the waveguide.  The
resonator causes the coupled pulse to be temporally
distorted\cite{HeebnerPRE02,TotsukaPRE07}, as shown in
Fig.~\ref{fig:theo_time}a.  The coherent addition of the coupled and uncoupled
pulses results in a time-varying polarization at the output.  When projecting
this polarization into the parallel or orthogonal axis, the detected peak is
either advanced or delayed in time with respect to the original one, illustrated
in Fig.~\ref{fig:theo_time}.%\cite{EPAPS_anim}.

The ratio between the amplitudes of the advanced and delayed polarizations (as
well as the corresponding group delays) can be adjusted by changing the coupling
between the resonator and the waveguide.  When the resonator is undercoupled,
the polarization conversion is weak and the output mostly conserves its original
polarization; the negative group delay is small, since the output is dominated
by the uncoupled pulse.  At critical coupling, both output polarizations are
equally transmitted, albeit with a total loss of half the input field. 
Operation in the strongly overcoupled regime maximizes the negative group delay
(temporal advancement) of the parallel polarization and reduces the overall
loss.  However, very strong polarization conversion\cite{BianucciOPL07}
significantly reduces the amplitude of the transmitted parallel polarization
while maximizing that of the orthogonal polarization.  To simplify the
measurement, we've chosen an intermediate case of a slightly overcoupled
fiber-resonator system.

Changing the input polarization will also result in a different ratio of
transmitted amplitudes (and group delays).  The largest (both positive and
negative) group delays are obtained for an input polarization of 45 degrees at
the expense of a somewhat reduced transmission amplitude for the parallel
polarized light.  For applications where a higher transmission of this component
is desirable, using an input polarization of $60$ degrees (as suggested in
Ref.~\cite{FietzOPL07b}) would increase this transmission at the expense of
smaller achievable group delays.  We choose an input polarization of $45$
degrees in our experiments to maximize the observable delays and bring them into
the range measurable with our existing instruments.

% Experimental results

\begin{figure}[htbp]
  \centering
  \includegraphics[width=8.2cm,keepaspectratio]{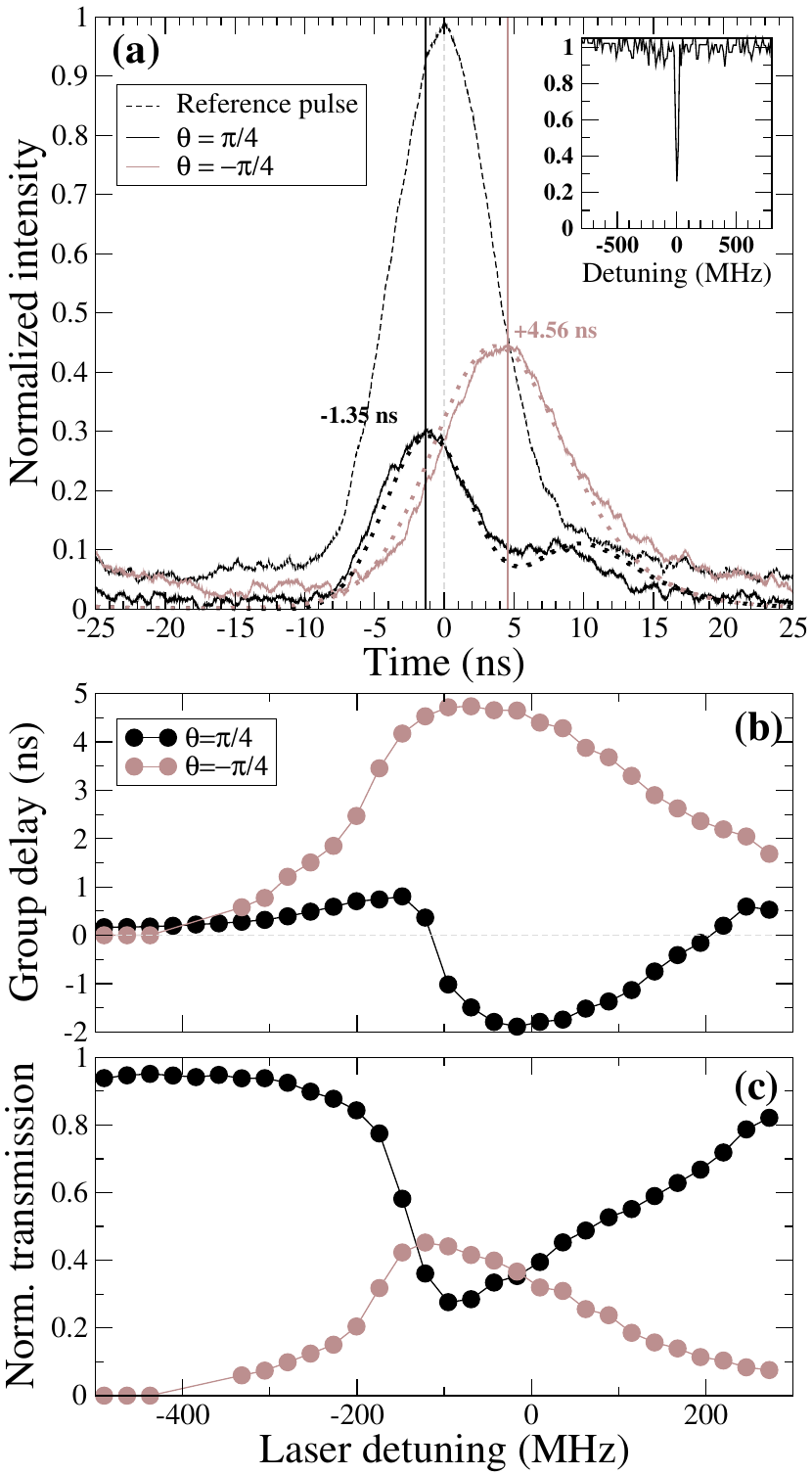}
  \caption{(a) Measured time traces showing negative and positive group delays
	polarization angles.  The dotted lines represent numerical calculations
	with estimated attenuation and coupling coefficients $a = 0.999998$ and
	$r = 0.999947$.  Inset:  Spectrum of the unloaded cavity resonance used
	in the experiments.  (b) Observed group delay measured for different
	laser frequencies.  (c) Experimental transmission spectra complementing
	the group delay data in panel b.}
  \label{fig:trace}
  \label{fig:delayvsf}
\end{figure}

Our experiments were done on a silica microsphere with a $49\,\micron$ diameter,
fabricated from a single-mode optical fiber using a $\mbox{CO}_2$ laser.  A
tapered optical fiber\cite{KnightOPL97,CaiPRL00} acted as the evanescently
coupled waveguide (the interaction length between the fiber and the sphere,
equivalent to the optical device length was estimated to be on the order of
$10\,\micron$).  The coupling strength was controlled by adjusting the relative
positions of the the sphere and fiber with a piezoelectric stage.  Using a
tunable diode laser we could find a reasonably sharp resonance, shown in the
inset of Fig.~\ref{fig:trace}a.  We estimate this resonance to have mode
numbers\cite{BohrenJWS83} $l \approx 465$ and $m \approx l$ and an unloaded Q
factor near $10^7$.

Setting the input polarization approximately at 45 degrees, we sent short
gaussian pulses ($\mbox{FWHM} \approx 8.4 \mbox{ns}$) generated by an
electro-optic modulator into the fiber.  We used a polarizer at the output of
the fiber to set the $\theta$ analyzer angle at $\pm\pi/4$, and a fast
photodiode connected to a digital oscilloscope to record time traces of the
field intensity.  Proper control of the polarization at the resonator and the
output was very important for the experiment, so we used two polarization
controllers to compensate for fiber birefringence, one before the resonator and
one after it.  When the resonator and fiber were brought closer together and the
incident light in resonance with the selected mode, the effected group delays
were noticeable in the measured traces.  Fig.~\ref{fig:trace}a shows traces for
both values of $\theta$ together with a reference trace taken with the resonator
uncoupled from the waveguide.  The pulse peak is advanced $1.35$ ns with respect
to the reference in the parallel trace, while it is delayed by $4.56$ ns in the
orthogonal trace.  The magnitudes of the changes in the arrival times for both
delayed and advanced pulses are much larger than what would be expected for as a
normal transit time through the device length.  From this we can infer that the
orthogonal polarization experienced a large group index causing a delay, while
for the parallel polarization the peak of the pulse left the resonator before
entering it (the corresponding group indices are $n_g^{\perp}\approx 10^5$
and $n_g^{\parallel}\approx -30000$).  Note that both the delayed and advanced
signals are of comparable intensities\cite{Coupling_note}.  Calculations using
the time-domain picture, including losses to higher order fiber modes of
$\approx 23\%$, show a reasonable agreement with the data. 

Both positive and negative group delays are intrinsically narrowband phenomena,
with a bandwidth given by that of the resonant mode.  As such, only pulses with
a carrier frequency matching that of the mode will be affected. 
Figure~\ref{fig:delayvsf}b corroborates this, showing that the delays become
smaller as the laser frequency sweeps away from the resonance center (the
corresponding transmissions are displayed in Fig.~\ref{fig:delayvsf}c).  Narrow
bandwidth is also responsible for some distortion of pulse shapes.  As with all
passive single-mode-based delay lines, the delay-bandwidth product is
limited.\cite{TuckerJLT05} Our high-Q resonators maximize the delay at the
expense of bandwidth, another choice intended to make the delays measurable with
our existing instruments, and thus our bandwidth is too narrow to allow for
applications at large modulation rates.  Besides, free-standing microspheres are
not easy to integrate on photonic circuits.  Other resonator systems, such as
silica microtoroids\cite{ArmaniNAT03} or silicon-on-insulator (SOI)
microrings\cite{LipsonOPM05} or silica could be better suited for integration. 
In particular, microfabricated SOI resonators have broader linewidths and could
work at higher modulation rates (at the expense of smaller induced group
delays).  Using chains of microfabricated resonators\cite{XiaNPH06,FietzOPL07b}
might be a way to sidestep that limitation.

\begin{figure}[htbp]
  \centering
  \includegraphics[width=8.2cm,keepaspectratio]{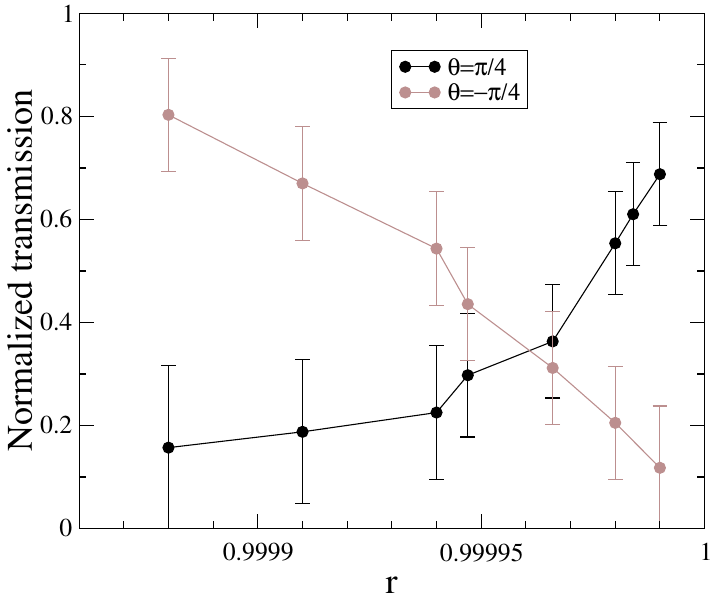}
  \caption{Experimental polarization conversion for different coupling
       strengths.  Significant conversion is observed as the coupling becomes
       stronger.}
  \label{fig:conv}
\end{figure}

The state of the polarization at the resonator is also important to obtain the
desired group delays\cite{FietzOPL07b}. The theoretical model predicts
significant conversion for the overcoupled system when the incoming polarization
is at 45 degrees\cite{BianucciOPL07}, so we use the presence of a strong
polarization conversion effect when the coupling increases as an indicator that
the polarization at the resonator is close to 45 degrees.  We can see in
Fig.~\ref{fig:conv} that the conversion becomes more pronounced as the coupling
increases (and $r$ decreases), giving us a confirmation that our incoming
polarization is acceptably close to the desired one.

% Conclusion

In summary, we have experimentally demonstrated that microresonator-loaded
optical fibers can be used to split an incident pulse into mutually orthogonal
output pulses experiencing positive and negative group delays.  The phenomenon
can be explained in both the spectral domain, through generalized Kramers-Kronig
relations and the time domain, via a resonator-induced polarization change. 
This capability could prove useful for the implementation of photonic circuits
in all-optical networks, for instance compensating processing-induced delays. 
Other applications such as single-photon devices for quantum computation and
communication are also within the realm of possibilities. 

\section{Acknowledgements}
We would like to thank Prof X.  Li for useful discussion and her technical
help.  This work was supported by NSF-NIRT (DMR-0210383), the Texas Advanced
Technology program, and the W.M.  Keck Foundation.  G.S.  and C.F.  acknowledge
support from ARO MURI grant no.  W911NF-04-01-0203.

% Bibliography


\begin{thebibliography}{28}
\expandafter\ifx\csname natexlab\endcsname\relax\def\natexlab#1{#1}\fi
\expandafter\ifx\csname bibnamefont\endcsname\relax
  \def\bibnamefont#1{#1}\fi
\expandafter\ifx\csname bibfnamefont\endcsname\relax
  \def\bibfnamefont#1{#1}\fi
\expandafter\ifx\csname citenamefont\endcsname\relax
  \def\citenamefont#1{#1}\fi
\expandafter\ifx\csname url\endcsname\relax
  \def\url#1{\texttt{#1}}\fi
\expandafter\ifx\csname urlprefix\endcsname\relax\def\urlprefix{URL }\fi
\providecommand{\bibinfo}[2]{#2}
\providecommand{\eprint}[2][]{\url{#2}}

\bibitem[{\citenamefont{Chang et~al.}(2006)\citenamefont{Chang, Yu, Yeo,
  Chowdhury, and Jia}}]{ChangPIEEE06}
\bibinfo{author}{\bibfnamefont{G.-K.} \bibnamefont{Chang}},
  \bibinfo{author}{\bibfnamefont{J.}~\bibnamefont{Yu}},
  \bibinfo{author}{\bibfnamefont{Y.-K.} \bibnamefont{Yeo}},
  \bibinfo{author}{\bibfnamefont{A.}~\bibnamefont{Chowdhury}},
  \bibnamefont{and} \bibinfo{author}{\bibfnamefont{Z.}~\bibnamefont{Jia}},
  \bibinfo{journal}{Proceedings of the IEEE} \textbf{\bibinfo{volume}{94}},
  \bibinfo{pages}{892} (\bibinfo{year}{2006}).

\bibitem[{\citenamefont{Hau et~al.}(1999)\citenamefont{Hau, Harris, Dutton, and
  Behroozi}}]{HauNAT99}
\bibinfo{author}{\bibfnamefont{L.~V.} \bibnamefont{Hau}},
  \bibinfo{author}{\bibfnamefont{S.~E.} \bibnamefont{Harris}},
  \bibinfo{author}{\bibfnamefont{Z.}~\bibnamefont{Dutton}}, \bibnamefont{and}
  \bibinfo{author}{\bibfnamefont{C.~H.} \bibnamefont{Behroozi}},
  \bibinfo{journal}{Nature} \textbf{\bibinfo{volume}{397}},
  \bibinfo{pages}{594} (\bibinfo{year}{1999}).

\bibitem[{\citenamefont{Wang et~al.}(2005)\citenamefont{Wang, Kuzmich, and
  Dogariu}}]{WangNature00}
\bibinfo{author}{\bibfnamefont{L.~J.} \bibnamefont{Wang}},
  \bibinfo{author}{\bibfnamefont{A.}~\bibnamefont{Kuzmich}}, \bibnamefont{and}
  \bibinfo{author}{\bibfnamefont{A.}~\bibnamefont{Dogariu}},
  \bibinfo{journal}{Nature} \textbf{\bibinfo{volume}{406}},
  \bibinfo{pages}{277} (\bibinfo{year}{2005}).

\bibitem[{\citenamefont{Camacho et~al.}(2007)\citenamefont{Camacho, Pack,
  Howell, Schweinsberg, and Boyd}}]{CamachoPRL07}
\bibinfo{author}{\bibfnamefont{R.~M.} \bibnamefont{Camacho}},
  \bibinfo{author}{\bibfnamefont{M.~V.} \bibnamefont{Pack}},
  \bibinfo{author}{\bibfnamefont{J.~C.} \bibnamefont{Howell}},
  \bibinfo{author}{\bibfnamefont{A.}~\bibnamefont{Schweinsberg}},
  \bibnamefont{and} \bibinfo{author}{\bibfnamefont{R.~W.} \bibnamefont{Boyd}},
  \bibinfo{journal}{Phys. Rev. Lett.} \textbf{\bibinfo{volume}{98}},
  \bibinfo{pages}{153601} (\bibinfo{year}{2007}).

\bibitem[{\citenamefont{Bigelow et~al.}(2003)\citenamefont{Bigelow, Lepeshkin,
  and Boyd}}]{BigelowSCI03}
\bibinfo{author}{\bibfnamefont{M.~S.} \bibnamefont{Bigelow}},
  \bibinfo{author}{\bibfnamefont{N.~N.} \bibnamefont{Lepeshkin}},
  \bibnamefont{and} \bibinfo{author}{\bibfnamefont{R.~W.} \bibnamefont{Boyd}},
  \bibinfo{journal}{Science} \textbf{\bibinfo{volume}{301}},
  \bibinfo{pages}{200} (\bibinfo{year}{2003}).

\bibitem[{\citenamefont{Chang-Hasnain and Chuang}(2006)}]{Chang-HasnainJLT06}
\bibinfo{author}{\bibfnamefont{C.~J.} \bibnamefont{Chang-Hasnain}}
  \bibnamefont{and} \bibinfo{author}{\bibfnamefont{S.~L.}
  \bibnamefont{Chuang}}, \bibinfo{journal}{J. Lightwave Technol.}
  \textbf{\bibinfo{volume}{24}}, \bibinfo{pages}{4642} (\bibinfo{year}{2006}).

\bibitem[{\citenamefont{Sarkar et~al.}(2006)\citenamefont{Sarkar, Guo, and
  Wang}}]{SarkarOPE06}
\bibinfo{author}{\bibfnamefont{S.}~\bibnamefont{Sarkar}},
  \bibinfo{author}{\bibfnamefont{Y.}~\bibnamefont{Guo}}, \bibnamefont{and}
  \bibinfo{author}{\bibfnamefont{H.}~\bibnamefont{Wang}},
  \bibinfo{journal}{Opt. Express} \textbf{\bibinfo{volume}{14}},
  \bibinfo{pages}{2845} (\bibinfo{year}{2006}).

\bibitem[{\citenamefont{Vlasov et~al.}(2005)\citenamefont{Vlasov, O'Boyle,
  Hamann, and McNab}}]{VlasovNAT05}
\bibinfo{author}{\bibfnamefont{Y.~A.} \bibnamefont{Vlasov}},
  \bibinfo{author}{\bibfnamefont{M.}~\bibnamefont{O'Boyle}},
  \bibinfo{author}{\bibfnamefont{H.~F.} \bibnamefont{Hamann}},
  \bibnamefont{and} \bibinfo{author}{\bibfnamefont{S.~J.} \bibnamefont{McNab}},
  \bibinfo{journal}{Nature} \textbf{\bibinfo{volume}{438}}, \bibinfo{pages}{65}
  (\bibinfo{year}{2005}).

\bibitem[{\citenamefont{Xia et~al.}(2007)\citenamefont{Xia, Sekaric, and
  Vlasov}}]{XiaNPH06}
\bibinfo{author}{\bibfnamefont{F.}~\bibnamefont{Xia}},
  \bibinfo{author}{\bibfnamefont{L.}~\bibnamefont{Sekaric}}, \bibnamefont{and}
  \bibinfo{author}{\bibfnamefont{Y.}~\bibnamefont{Vlasov}},
  \bibinfo{journal}{Nat. Photon.} \textbf{\bibinfo{volume}{1}},
  \bibinfo{pages}{65} (\bibinfo{year}{2007}).

\bibitem[{\citenamefont{Totsuka and Tomita}(2007)}]{TotsukaPRE07}
\bibinfo{author}{\bibfnamefont{K.}~\bibnamefont{Totsuka}} \bibnamefont{and}
  \bibinfo{author}{\bibfnamefont{M.}~\bibnamefont{Tomita}},
  \bibinfo{journal}{Phys. Rev. E} \textbf{\bibinfo{volume}{75}},
  \bibinfo{pages}{016610} (\bibinfo{year}{2007}).

\bibitem[{\citenamefont{Totsuka et~al.}(2007)\citenamefont{Totsuka, Kobayashi,
  and Tomita}}]{TotsukaPRL07}
\bibinfo{author}{\bibfnamefont{K.}~\bibnamefont{Totsuka}},
  \bibinfo{author}{\bibfnamefont{N.}~\bibnamefont{Kobayashi}},
  \bibnamefont{and} \bibinfo{author}{\bibfnamefont{M.}~\bibnamefont{Tomita}},
  \bibinfo{journal}{Phys. Rev. Lett.} \textbf{\bibinfo{volume}{98}},
  \bibinfo{pages}{213904} (\bibinfo{year}{2007}).

\bibitem[{\citenamefont{Solli et~al.}(2002)\citenamefont{Solli, Chiao, and
  Hickmann}}]{SolliPRE02}
\bibinfo{author}{\bibfnamefont{D.}~\bibnamefont{Solli}},
  \bibinfo{author}{\bibfnamefont{R.~Y.} \bibnamefont{Chiao}}, \bibnamefont{and}
  \bibinfo{author}{\bibfnamefont{J.~M.} \bibnamefont{Hickmann}},
  \bibinfo{journal}{Phys.~Rev.~E} \textbf{\bibinfo{volume}{66}},
  \bibinfo{pages}{056601} (\bibinfo{year}{2002}).

\bibitem[{\citenamefont{Knight et~al.}(1997)\citenamefont{Knight, Cheung,
  Jacques, and Birks}}]{KnightOPL97}
\bibinfo{author}{\bibfnamefont{J.~C.} \bibnamefont{Knight}},
  \bibinfo{author}{\bibfnamefont{G.}~\bibnamefont{Cheung}},
  \bibinfo{author}{\bibfnamefont{F.}~\bibnamefont{Jacques}}, \bibnamefont{and}
  \bibinfo{author}{\bibfnamefont{T.~A.} \bibnamefont{Birks}},
  \bibinfo{journal}{Opt. Lett.} \textbf{\bibinfo{volume}{22}},
  \bibinfo{pages}{1129} (\bibinfo{year}{1997}).

\bibitem[{\citenamefont{Cai et~al.}(2000)\citenamefont{Cai, Painter, and
  Vahala}}]{CaiPRL00}
\bibinfo{author}{\bibfnamefont{M.}~\bibnamefont{Cai}},
  \bibinfo{author}{\bibfnamefont{O.}~\bibnamefont{Painter}}, \bibnamefont{and}
  \bibinfo{author}{\bibfnamefont{K.~J.} \bibnamefont{Vahala}},
  \bibinfo{journal}{Phys. Rev. Lett.} \textbf{\bibinfo{volume}{85}},
  \bibinfo{pages}{74} (\bibinfo{year}{2000}).

\bibitem[{\citenamefont{Boyd and Gauthier}(2002)}]{BoydPIP02}
\bibinfo{author}{\bibfnamefont{R.~W.} \bibnamefont{Boyd}} \bibnamefont{and}
  \bibinfo{author}{\bibfnamefont{D.~J.} \bibnamefont{Gauthier}},
  \bibinfo{journal}{Prog. Optics} \textbf{\bibinfo{volume}{43}},
  \bibinfo{pages}{497} (\bibinfo{year}{2002}).

\bibitem[{\citenamefont{Solli et~al.}(2003)\citenamefont{Solli, McCormick,
  Ropers, Morehead, Chiao, and Hickmann}}]{SolliPRL03}
\bibinfo{author}{\bibfnamefont{D.~R.} \bibnamefont{Solli}},
  \bibinfo{author}{\bibfnamefont{C.~F.} \bibnamefont{McCormick}},
  \bibinfo{author}{\bibfnamefont{C.}~\bibnamefont{Ropers}},
  \bibinfo{author}{\bibfnamefont{J.~J.} \bibnamefont{Morehead}},
  \bibinfo{author}{\bibfnamefont{R.~Y.} \bibnamefont{Chiao}}, \bibnamefont{and}
  \bibinfo{author}{\bibfnamefont{J.~M.} \bibnamefont{Hickmann}},
  \bibinfo{journal}{Phys. Rev. Lett.} \textbf{\bibinfo{volume}{91}},
  \bibinfo{pages}{143906} (\bibinfo{year}{2003}).

\bibitem[{\citenamefont{Solli et~al.}(2004)\citenamefont{Solli, McCormick,
  Chiao, Popescu, and Hickmann}}]{SolliPRL04}
\bibinfo{author}{\bibfnamefont{D.~R.} \bibnamefont{Solli}},
  \bibinfo{author}{\bibfnamefont{C.~F.} \bibnamefont{McCormick}},
  \bibinfo{author}{\bibfnamefont{R.~Y.} \bibnamefont{Chiao}},
  \bibinfo{author}{\bibfnamefont{S.}~\bibnamefont{Popescu}}, \bibnamefont{and}
  \bibinfo{author}{\bibfnamefont{J.~M.} \bibnamefont{Hickmann}},
  \bibinfo{journal}{Phys. Rev. Lett.} \textbf{\bibinfo{volume}{92}},
  \bibinfo{pages}{043601} (\bibinfo{year}{2004}).

\bibitem[{\citenamefont{Melloni and Morichetti}(2007)}]{MelloniPRL07}
\bibinfo{author}{\bibfnamefont{A.}~\bibnamefont{Melloni}} \bibnamefont{and}
  \bibinfo{author}{\bibfnamefont{F.}~\bibnamefont{Morichetti}},
  \bibinfo{journal}{Phys. Rev. Lett.} \textbf{\bibinfo{volume}{98}},
  \bibinfo{pages}{173902} (\bibinfo{year}{2007}).

\bibitem[{\citenamefont{Bianucci et~al.}(2007)\citenamefont{Bianucci, Fietz,
  Robertson, Shvets, and Shih}}]{BianucciOPL07}
\bibinfo{author}{\bibfnamefont{P.}~\bibnamefont{Bianucci}},
  \bibinfo{author}{\bibfnamefont{C.~R.} \bibnamefont{Fietz}},
  \bibinfo{author}{\bibfnamefont{J.~W.} \bibnamefont{Robertson}},
  \bibinfo{author}{\bibfnamefont{G.}~\bibnamefont{Shvets}}, \bibnamefont{and}
  \bibinfo{author}{\bibfnamefont{C.-K.} \bibnamefont{Shih}},
  \bibinfo{journal}{Opt. Lett.} \textbf{\bibinfo{volume}{32}},
  \bibinfo{pages}{2224} (\bibinfo{year}{2007}).

\bibitem[{\citenamefont{Fietz and Shvets}(2007{\natexlab{a}})}]{FietzOPL07}
\bibinfo{author}{\bibfnamefont{C.}~\bibnamefont{Fietz}} \bibnamefont{and}
  \bibinfo{author}{\bibfnamefont{G.}~\bibnamefont{Shvets}},
  \bibinfo{journal}{Opt. Lett.} \textbf{\bibinfo{volume}{32}},
  \bibinfo{pages}{1683} (\bibinfo{year}{2007}{\natexlab{a}}).

\bibitem[{\citenamefont{Heebner et~al.}(2002)\citenamefont{Heebner, Boyd, and
  Park}}]{HeebnerPRE02}
\bibinfo{author}{\bibfnamefont{J.~E.} \bibnamefont{Heebner}},
  \bibinfo{author}{\bibfnamefont{R.~W.} \bibnamefont{Boyd}}, \bibnamefont{and}
  \bibinfo{author}{\bibfnamefont{Q.-H.} \bibnamefont{Park}},
  \bibinfo{journal}{Phys. Rev. E} \textbf{\bibinfo{volume}{65}},
  \bibinfo{pages}{036619} (\bibinfo{year}{2002}).

%\bibitem[{EPA()}]{EPAPS_anim}
%\bibinfo{note}{See EPAPS Document No. [xxx] for a visual illustration of the
%  temporal behaviour of both polarization components. For more information on
%  EPAPS, see http://www.aip.org/pubservs/epaps.html.}

\bibitem[{\citenamefont{Fietz and Shvets}(2007{\natexlab{b}})}]{FietzOPL07b}
\bibinfo{author}{\bibfnamefont{C.}~\bibnamefont{Fietz}} \bibnamefont{and}
  \bibinfo{author}{\bibfnamefont{G.}~\bibnamefont{Shvets}},
  \bibinfo{journal}{Opt. Lett.} \textbf{\bibinfo{volume}{32}},
  \bibinfo{pages}{3480} (\bibinfo{year}{2007}{\natexlab{b}}).

\bibitem[{\citenamefont{Bohren and Huffman}(1983)}]{BohrenJWS83}
\bibinfo{author}{\bibfnamefont{C.~F.} \bibnamefont{Bohren}} \bibnamefont{and}
  \bibinfo{author}{\bibfnamefont{D.~R.} \bibnamefont{Huffman}},
  \emph{\bibinfo{title}{Absorption and Scattering of Light by Small Particles}}
  (\bibinfo{publisher}{John Wiley \& Sons}, \bibinfo{address}{New York},
  \bibinfo{year}{1983}).

\bibitem[{Cou()}]{Coupling_note}
\bibinfo{note}{The fact that the orthogonal peak is smowhat stronger than the
  parallel one lets us infer that the resonator is already into the overcoupled
  regime, but not far away from critical coupling.}

\bibitem[{\citenamefont{Tucker et~al.}(2005)\citenamefont{Tucker, Ku, and
  Chang-Hasnain}}]{TuckerJLT05}
\bibinfo{author}{\bibfnamefont{R.}~\bibnamefont{Tucker}},
  \bibinfo{author}{\bibfnamefont{P.-C.} \bibnamefont{Ku}}, \bibnamefont{and}
  \bibinfo{author}{\bibfnamefont{C.}~\bibnamefont{Chang-Hasnain}},
  \bibinfo{journal}{J. Lightwave Technol.} \textbf{\bibinfo{volume}{23}},
  \bibinfo{pages}{4046} (\bibinfo{year}{2005}).

\bibitem[{\citenamefont{Armani et~al.}(2003)\citenamefont{Armani, Kippenberg,
  Spillane, and Vahala}}]{ArmaniNAT03}
\bibinfo{author}{\bibfnamefont{D.~K.} \bibnamefont{Armani}},
  \bibinfo{author}{\bibfnamefont{T.~J.} \bibnamefont{Kippenberg}},
  \bibinfo{author}{\bibfnamefont{S.~M.} \bibnamefont{Spillane}},
  \bibnamefont{and} \bibinfo{author}{\bibfnamefont{K.~J.}
  \bibnamefont{Vahala}}, \bibinfo{journal}{Nature (London)}
  \textbf{\bibinfo{volume}{421}}, \bibinfo{pages}{925} (\bibinfo{year}{2003}).

\bibitem[{\citenamefont{Lipson}(2005)}]{LipsonOPM05}
\bibinfo{author}{\bibfnamefont{M.}~\bibnamefont{Lipson}},
  \bibinfo{journal}{Opt. Mater.} \textbf{\bibinfo{volume}{27}},
  \bibinfo{pages}{731} (\bibinfo{year}{2005}).

\end{thebibliography}
\end{document}